\documentclass[10pt,aps,prd,showpacs,twocolumn,superscriptaddress,longbibliography,nofootinbib]{revtex4-1}
\usepackage[english]{babel}
\usepackage{amsfonts}
\usepackage{amssymb}
\usepackage{amsmath}
\usepackage{amsthm}
\usepackage{mathrsfs}
\usepackage{graphicx}
\usepackage{upgreek}
\usepackage{euscript}
\usepackage{color,soul,ulem}
\usepackage{ulem}
\usepackage{natbib}

\usepackage{etoolbox}
\apptocmd{\sloppy}{\hbadness 10000\relax}{}{}

\usepackage{mathtools}

\let\originalleft\left
\let\originalright\right
\renewcommand{\left}{\mathopen{}\mathclose\bgroup\originalleft}
\renewcommand{\right}{\aftergroup\egroup\originalright}

\makeatletter
\newenvironment{equations}[1][]{\subequations\ifx\relax#1\relax\else\label{#1}
\fi\align\ignorespaces}{\endalign\ignorespacesafterend\endsubequations}
\def\@spliteq#1{\begin{equation}\begin{split}#1\end{split}\end{equation}}
\def\splitequation{\collect@body\@spliteq}

\makeatother

\date{\today}

\newcommand{\eqend}[1]{\,#1}
\newcommand{\pp}[1]{\left( #1 \right)}
\newcommand{\cc}[1]{\left[ #1 \right]}
\newcommand{\ch}[1]{\left\{ #1 \right\}}
\newcommand{\tr}{\textrm{tr}}
\newcommand{\del}{\partial}
\newcommand{\vt}[1]{\textbf{\textrm{#1}}}

\begin{document}

\title{Equivalence between the in-in perturbation theories for quantum fields in Minkowski spacetime and in the Rindler wedge}

\author{Atsushi Higuchi}
\email{atsushi.higuchi@york.ac.uk}
\affiliation{Department of Mathematics, University of York, Heslington, York YO10 5DD, United Kingdom}

\author{William~C.~C.~Lima}
\altaffiliation{\textit{Current address}: Department of Mathematics, University of York, Heslington, York YO10 5DD, United Kingdom}
\email{william.correadelima@york.ac.uk}
\affiliation{Centro de Ci\^encias Naturais e Humanas, Universidade Federal do ABC, Avenida dos Estados, 5001, Bang\'u, 09210-580, Santo Andr\'e, S\~ao Paulo, Brazil}


\begin{abstract}
We investigate the relation between the time-ordered vacuum correlation functions for interacting real scalar fields in Minkowski spacetime and in the Rindler wedge. The correlation functions are constructed perturbatively within the in-in formalism, often employed in calculations in more general spacetimes. We prove to all orders in perturbation theory that the time-ordered vacuum correlation functions can be calculated in the in-in formalism with internal vertices restricted to any Rindler wedge containing the external points. This implies that the Minkowski in-in (or in-out) perturbative expansion of the vacuum correlation functions is reproduced by the Rindler in-in perturbative expansion of these correlators in a thermal state at the Unruh temperature.
\end{abstract}

\maketitle

\section{Introduction}                                                                                                                               %
\label{sec:introduction}                                                                                                                             %

\begin{center}
 \begin{figure}[ht]
   \includegraphics[scale=0.6]{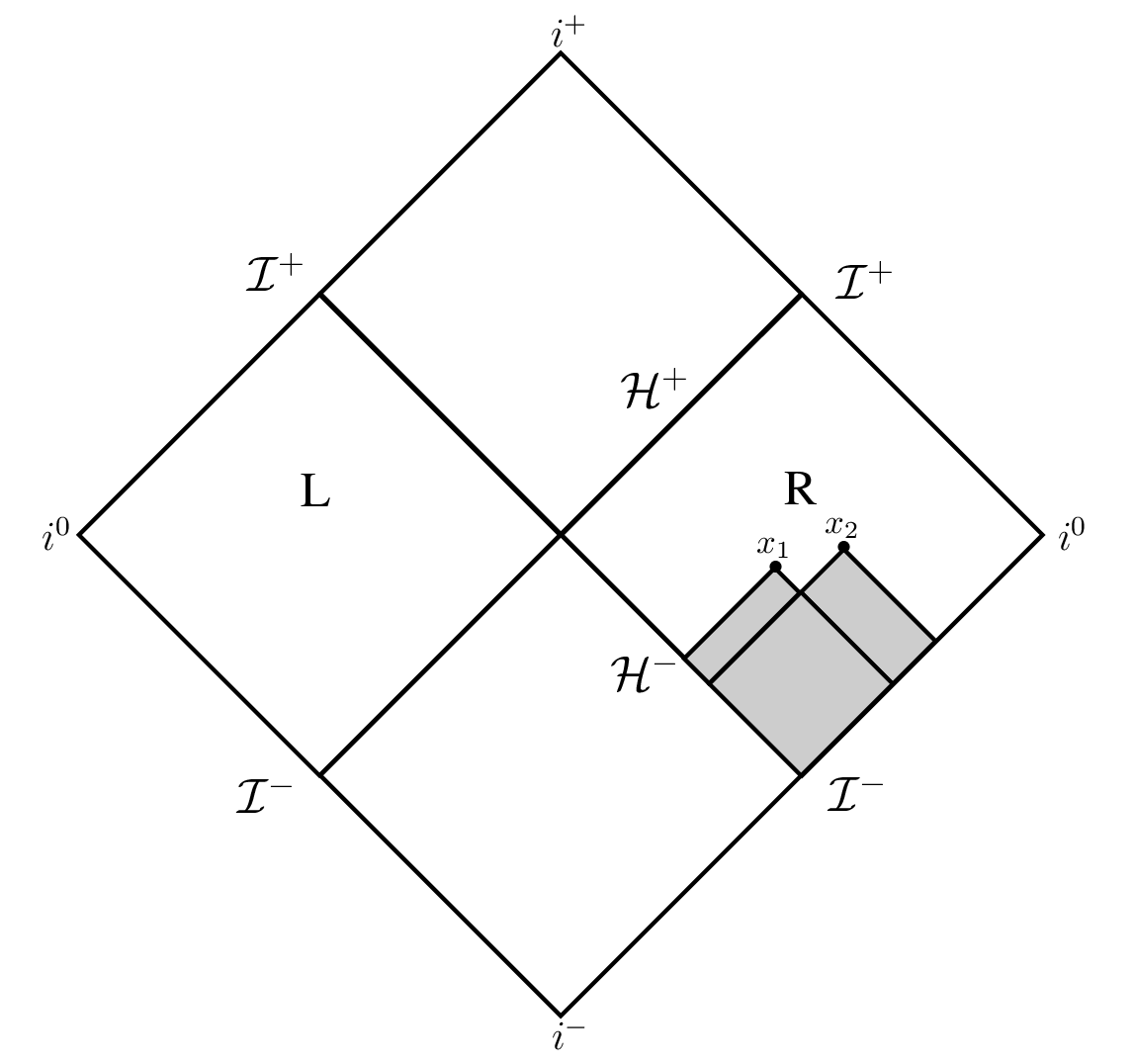}
   \caption{A conformal diagram of Minkowski spacetime. The L and R regions are the left and right Rindler wedges, respectively. The gray region
            depicts the portion of the past light cones emanating from the spacetime points $x_1$ and $x_2$ inside the right wedge, while the null
            surfaces $\mathcal{H}^\pm$ are the future and past Rindler horizons, respectively.}
   \label{fig:minkowski_spacetime}
 \end{figure}
\end{center}

It is well known that uniformly accelerated observers in Minkowski spacetime with proper acceleration $a$ perceive the vacuum of a quantum field as a thermal equilibrium state at the temperature $T_\textrm{U} \equiv a/2\pi$ (the Unruh effect~\cite{unruh_prd_1976}). The accelerated observers follow the integral curves of the timelike Killing vector field that generates Lorentz boosts, and thus are restricted to the regions~L or R in Fig.~\ref{fig:minkowski_spacetime}, the left or right Rindler wedge. Clearly, inertial and accelerated observers will have quite different descriptions for the same physical phenomenon, e.g., in the case of the decay of accelerated particles~\cite{matsas_vanzella_prd_1999, vanzella_matsas_prl_2001}. Although these two descriptions are clearly different, they are completely equivalent when it comes to the prediction of observables, as the consistency of the theory demands~\cite{crispino_higuchi_matsas_rmp_2008}.

Indeed, Bisognano and Wichmann~\cite{bisognano_wichmann_jmp_1975} proved a theorem showing that the vacuum expectation value of any observable
supported inside the Rindler wedge corresponds to a statistical average in a KMS state~\cite{kubo_jpsj_1957, martin_schwinger_pr_1959,
haag_hugenholtz_Winnink_cmp_1967} at the Unruh temperature with respect to the generator of the relevant Lorentz boosts. Their result was obtained in
the axiomatic approach~\cite{streater_wightman_book} and is valid for all interacting quantum field theories satisfying the Wightman axioms. The
importance of the Bisognano-Wichmann theorem to the Unruh effect was only recognized later by Sewell~\cite{sewell_ap_1982}, who obtained a similar
result for more general spacetimes. The equivalence between the inertial and accelerated pictures can also be verified via the Euclidean theory,
with the Cartesian and cylindrical coordinates for Euclidean space corresponding to the Minkowski and Rindler spacetimes respectively (see below),
as the analytic continuation of the $N$-point correlation functions from imaginary to real times is known to define a unique
state~\cite{osterwalder_schrader_cmp_1973, osterwalder_schrader_cmp_1975}.

Notwithstanding the importance of the general results described above, in the application of quantum field theory to realistic problems one often
relies on perturbative methods to treat interacting models. Therefore, it is important to analyze how the equivalence between the inertial and
accelerated descriptions emerge from a perturbative computation. In the Euclidean formulation, one constructs the $N$-point correlation functions by
first performing the vertex integrations on the entire Euclidean space and then the analytic continuation of the result to real (inertial or
Rindler) time. The inertial coordinates correspond to Cartesian coordinates of Euclidean space while the Rindler coordinates correspond to cylindrical
coordinates, with the imaginary part of the Rindler time as the angular coordinate. Hence, the Euclidean correlators define both the vacuum state in Minkowski spacetime and a KMS state in the Rindler wedge, for both free~\cite{sciama_candelas_deutsch_ap_1981, fulling_ruijsenaars_pr_1987} and
interacting~\cite{gibbons_perry_prl_1976, gibbons_perry_prsl_1978} theories. This can readily be seen from the path-integral formulation of the problem, as discussed by Unruh and Weiss~\cite{unruh_weiss_prd_1984}.

Although the perturbative Euclidean formalism is largely employed in practical computations in thermal field theory, there is a Lorentzian version of
it known as the Schwinger-Keldysh formalism~\cite{schwinger_jmp_1961, keldysh_zetf_1964}, often used in more general spacetimes. This formalism agrees
with the so-called in-in (also known as closed-time path or real-time) formalism of quantum field theory if the condition called factorization
is satisfied~\cite{landsman_van_weert_pr_1987}. In the in-in formalism, one computes expectation values and correlation functions rather than transition
matrix elements~\cite{chou_et_al_pr_1985}. It is well known that the in-in formalism is causal: all the vertex integrals in the diagrammatic expansion can be restricted to the union of the past light cones emanating from the external points.

Since the Euclidean formalism agrees with both the in-in formalism in Minkowski spacetime and Schwinger-Keldysh formalism (which is equivalent to the in-in formalism if the factorization property holds) in the Rindler wedge, the verification of the factorization property in the Rindler wedge would imply the equivalence between the in-in perturbation theories in Minkowski spacetime and the Rindler wedge. The physical content of the factorization property is that one can take the free-theory thermal state as the initial state of the interacting system in the limit where the initial time lies far in the past. In Minkowski thermal field theory, the factorization is due to the decay of the free-field propagator at large timelike distances. This property has also been verified for self-interacting massive scalar fields in the static patch of de~Sitter spacetime at all orders in perturbation theory~\cite{higuchi_marolf_morrison_prd_2011}. This was a key step in establishing the equivalence of the Euclidean and interacting Bunch-Davies vacua for massive scalar fields in the Poincar\'e patch of de~Sitter spacetime. In the de~Sitter case, the proof of the factorization property is facilitated by the fact that in the
static patch the spatial section has finite volume~\cite{higuchi_marolf_morrison_prd_2011}. In the Rindler wedge the volume of the Rindler spatial section is not finite, posing some difficulties in verifying the factorization property unlike in the de~Sitter case.

Rather than attempting to prove factorization directly, we adopt the following strategy for showing the equivalence between the in-in perturbation theories in the whole of Minkowski spacetime and in the Rindler wedge for a self-interacting massive real scalar field (with non-derivative interactions) in the Minkowski vacuum state. We start from the perturbative expansion of the time-ordered $N$-point correlation functions as defined by the in-in formalism in $n$-dimensional Minkowski spacetime, noting that the in-in formalism is equivalent to the usual in-out formalism in this spacetime. We
then prove that if all the $N$ external points of a diagram are inside the Rindler wedge, then the contribution coming from the integration of any of
its internal vertices outside the Rindler wedge cancel out, under the assumption that the one-point function of the scalar field vanishes. This result will imply that the vacuum correlation functions in Minkowski spacetime are equivalent to the thermal correlation functions in the Rindler wedge at temperature $T_\textrm{U}$ \textit{to all orders} in the in-in formalism, as stated in Ref.~\cite{higuchi_marolf_morrison_prd_2011} without proof. It is interesting to note that, due to the symmetries of the Minkowski vacuum, our result holds irrespective of where we place the Rindler horizons, as long as all the external points fall within the wedge.

The rest of the paper is organized as follows. We first review the Schwinger-Keldysh formalism in a general static spacetime background, which includes the Rindler wedge, (with possible external static fields) in Sec.~\ref{sec:in_in_formalism}. We then move on to the analysis of the perturbative expansion of the time-ordered $N$-point functions of a massive, self-interacting real scalar field theory in the Minkowski vacuum. In Sec.~\ref{sec:equivalence_rindler_minkowski} we prove that these $N$-point functions can be computed in the Rindler wedge containing the external points in the in-in formalism, under the assumption that the one-point function of the scalar field vanishes. In other words, we show that the integral for the vertices can be restricted to this Rindler wedge. (This calculation is known to correspond to the in-in formalism as defined by accelerated observers in a KMS state at the Unruh temperature with respect to their proper time.) To establish this fact, we first prove that time-ordered $N$-point functions can be computed in the region between two parallel null planes containing the external points using either the in-in or in-out formalism. This result is illustrated by an explicit computation of a particular diagram in Sec.~\ref{sec:explicit_example}. We conclude in Sec.~\ref{sec:discussion} with a discussion of our results. A brief account of the light-cone quantization of a free scalar field can be found in Appendix~\ref{sec:appendix_light_cone_quantisation}. Throughout this paper we employ units such that $k_\textrm{B} = \hbar = c = 1$ and adopt the signature $(-++\dots +)$ for the metric.

\section{Preliminaries}                                                                                                                              %
\label{sec:in_in_formalism}                                                                                                                          %

\subsection{The Schwinger-Keldysh formalism}\label{sec:schwinger-keldysh}
We consider a static spacetime and let $H$ denote the Hamiltonian operator of the quantum system under consideration. We assume that it has the form
\begin{equation}\label{eq:H}
 H = H_0 + V\eqend{,}
\end{equation}
where $H_0$ is the Hamiltonian operator of the free system and $V$ is the interaction term. Both $H_0$ and $V$ are assumed to be Hermitian. We let $U_\textrm{H}(t,t_\textrm{i})= e^{-iH(t-t_\textrm{i})}$ be the evolution operator from the initial time $t_\textrm{i}$ in the Heisenberg picture and define the evolution operator in the interaction picture as
\begin{equation}\label{eq:U_I}
 U_\textrm{I}(t,t_\textrm{i}) \equiv e^{iH_0(t - t_\textrm{i})}U_\textrm{H}(t,t_\textrm{i})\eqend{.}
\end{equation}
The operator $U_\textrm{I}(t,t_\textrm{i})$
satisfies the equation of motion
\begin{equation}\label{eq:U_I_eom}
 i\frac{d}{dt}U_\textrm{I}(t,t_\textrm{i}) = H_\textrm{I}(t) U_\textrm{I}(t,t_\textrm{i})\eqend{,}
\end{equation}
with the initial condition $U_\textrm{I}(t_\textrm{i},t_\textrm{i}) = I$, where $I$ is the identity operator. The interaction Hamiltonian operator 
$H_\textrm{I}(t)$ is defined by
\begin{equation}\label{eq:H_I}
 H_\textrm{I}(t) \equiv e^{iH_0 (t - t_\textrm{i})} V e^{-iH_0 (t - t_\textrm{i})}\eqend{.}
\end{equation}

The operator $U_\textrm{I}(t,t_\textrm{i})$ is unitary as can be seen from Eq.~(\ref{eq:U_I}). We define the operators $U_\textrm{I}(t,t')$ for arbitrary values of $t$ and $t'$ by letting
\begin{equation}\label{eq:U_I_composition_dagger}
 U_\textrm{I}(t,t') = U_\textrm{I}(t,t_\textrm{i})U_\textrm{I}(t',t_\textrm{i})^\dagger\eqend{.}
\end{equation}
The operators $U_\textrm{I}(t,t')$ can be written in terms of Dyson's series~\cite{dyson_pr_1949}:
\begin{equation}\label{eq:dyson_series}
 U_\textrm{I}(t,t') = T\exp\ch{-i\int_{t'}^tdt''H_\textrm{I}(t'')},\,\, t \geq t'\eqend{,}
\end{equation}
where $T$ indicates the time-ordering. The evolution operator $U_\textrm{I}(t,t')$ has the following useful property:
\begin{equation}\label{eq:U_I_composition}
 U_\textrm{I}(t_3,t_2)U_\textrm{I}(t_2,t_1) =  U_\textrm{I}(t_3,t_1)\eqend{.}
\end{equation}

Here we are interested in computing the correlation functions of the observables in a certain state. In the Heisenberg picture, the time evolution of an
observable $A$ is given by 
\begin{equation}\label{eq:evolution_A_H}
 A_\textrm{H}(t) \equiv U_\textrm{H}(t,t_\textrm{i})^\dagger A U_\textrm{H}(t,t_\textrm{i})\eqend{,}
\end{equation}
with the initial condition $A_\textrm{H}(t_\textrm{i}) = A$. In general, the state of the system is described by a (normalized) density matrix $\rho$.
The time-ordered correlation function in the state $\rho$ of, e.g., two observables $A$ and $B$ at different times is then defined by the following
trace over the Hilbert space of states:
\begin{equation}\label{eq:time_ordered_correlation}
 \langle T[A(t_1)B(t_2)]\rangle \equiv \tr\{\rho\, T[A_\textrm{H}(t_1)B_\textrm{H}(t_2)]\}.
\end{equation}
The time evolution of the observables in the interaction picture is determined by the free Hamiltonian operator as
\begin{equation}\label{eq:evolution_A_I}
 A_\textrm{I}(t) \equiv e^{iH_0(t - t_\textrm{i})} A e^{-iH_0(t - t_\textrm{i})}\eqend{.}
\end{equation}
This operator is related to the Heisenberg operator $A_\textrm{H}(t)$ through Eq.~(\ref{eq:evolution_A_H}) as
\begin{equation}\label{eq:A_H_A_I_relation}
 A_\textrm{H}(t) = U_\textrm{I}(t_\textrm{i},t)A_\textrm{I}(t)U_\textrm{I}(t,t_\textrm{i})\eqend{.}
\end{equation}
Using Eq.~(\ref{eq:A_H_A_I_relation}), we can express the right-hand side of Eq.~(\ref{eq:time_ordered_correlation}) entirely in terms of
interaction-picture operators as
\begin{splitequation}\label{eq:time_ordered_correlation_interaction_pic}
\langle T[A(t_1)B(t_2)]\rangle & = \tr\{\rho\, U_\textrm{I}(t_\textrm{i},t_1)A_\textrm{I}(t_1)U_\textrm{I}(t_1,t_2) \\
                               &\ \times B_\textrm{I}(t_2) U_\textrm{I}(t_2,t_\textrm{i})\},\ \  t_1 > t_2 > t_\textrm{i}\eqend{.}
\end{splitequation}
If $t_2 > t_1> t_\textrm{i}$, then $\langle T[A(t_1)B(t_2)]\rangle$ is obtained from Eq.~(\ref{eq:time_ordered_correlation_interaction_pic}) by letting 
$t_1 \leftrightarrow t_2$ and $A_\textrm{I} \leftrightarrow B_\textrm{I}$.

We now consider a system in thermal equilibrium at inverse temperature $\beta$, i.e.
\begin{equation}\label{eq:interacing_thermal_state}
 \rho = \frac{e^{-\beta H}}{Z(\beta)}\eqend{,}
\end{equation}
where $Z(\beta) \equiv \tr\, e^{-\beta H}$ is the partition function. The operator on the right-hand side of Eq.~(\ref{eq:interacing_thermal_state}) can
be seen as an evolution operator in the imaginary time $-i\beta$ and conveniently expressed as~\cite{matsubara_ptp_1955}
\begin{equation}\label{eq:interaction_pic_thermal_state}
 e^{-\beta H} = e^{-\beta H_0}U_\textrm{I}(t_\textrm{i} - i\beta,t_\textrm{i})\eqend{.}
\end{equation}

We use Eqs.~(\ref{eq:interacing_thermal_state}) and~(\ref{eq:interaction_pic_thermal_state}) to write the correlator (\ref{eq:time_ordered_correlation_interaction_pic}) as
\begin{splitequation}\label{eq:time_ordered_correlation_interaction_pic_i_beta}
\langle T[A(t_1)B(t_2)]\rangle & = \frac{1}{Z(\beta)}\tr\{e^{-\beta H_0} U_\textrm{I}(t_\textrm{i} - i\beta,t_\textrm{i})
                                   U_\textrm{I}(t_\textrm{i},t_1) \\
                               &\ \times A_\textrm{I}(t_1) U_\textrm{I}(t_1,t_2) B_\textrm{I}(t_2) U_\textrm{I}(t_2,t_\textrm{i})\}\eqend{,}
\end{splitequation}
if $t_1 > t_2 > t_\textrm{i}$. In order to express Eq.~(\ref{eq:time_ordered_correlation_interaction_pic_i_beta}) in terms of the free thermal state
$\rho_0 = e^{-\beta H_0}/Z_0(\beta)$, we again employ Eq.~(\ref{eq:interaction_pic_thermal_state}) to obtain the following form for the interacting
partition function:
\begin{equation}\label{eq:partition_function_interaction_pic}
 Z(\beta) = Z_0(\beta)\tr\{\rho_0 U_\textrm{I}(t_\textrm{i} - i\beta,t_\textrm{i})\}\eqend{.}
\end{equation}
From Eqs.~(\ref{eq:time_ordered_correlation_interaction_pic_i_beta}) and~(\ref{eq:partition_function_interaction_pic}) we find
\begin{splitequation}\label{eq:correlator_interaction_pic}
 &\langle T[A(t_1)B(t_2)]\rangle \\
 &\, = \frac{\tr\{\rho_0 U_\textrm{I}(t_\textrm{i} - i\beta, t_\textrm{i})U_\textrm{I}(t_\textrm{i}, t_\textrm{f})
 T[U_\textrm{I}(t_\textrm{f}, t_\textrm{i})A_\textrm{I}(t_1)B_\textrm{I}(t_2)]\}}{\tr\{\rho_0
U_\textrm{I}(t_\textrm{i} - i\beta, t_\textrm{i})U_\textrm{I}(t_\textrm{i}, t_\textrm{f})U_\textrm{I}(t_\textrm{f}, t_\textrm{i})\}}\eqend{.}
\end{splitequation}
In obtaining this equality we have introduced $t_\textrm{f} > t_1, t_2$ and made use of property~(\ref{eq:U_I_composition}). We note that
\begin{splitequation}
 &T[U_\textrm{I}(t_\textrm{f}, t_\textrm{i})A_\textrm{I}(t_1)B_\textrm{I}(t_2)]\\
 &\; = \left\{
 \begin{array}{ll}
 U_\textrm{I}(t_\textrm{f},t_1)A_\textrm{I}(t_1)U_\textrm{I}(t_1,t_2)B_\textrm{I}(t_2)U_\textrm{I}(t_2,t_\textrm{i}) & \textrm{if}\,\, t_1 > t_2\eqend{,}\\
 U_\textrm{I}(t_\textrm{f},t_2)B_\textrm{I}(t_2)U_\textrm{I}(t_2,t_1)A_\textrm{I}(t_1)U_\textrm{I}(t_1,t_\textrm{i}) & \textrm{if}\,\, t_2 > t_1\eqend{.}
 \end{array}
 \right.
\end{splitequation}
The right-hand side of Eq.~(\ref{eq:correlator_interaction_pic}) is independent of $t_\textrm{f}$ as long as it is larger than both $t_1$ and $t_2$. Thus, the Schwinger-Keldysh formalism is causal.

\begin{center}
 \begin{figure}[ht]
   \includegraphics[scale=0.6]{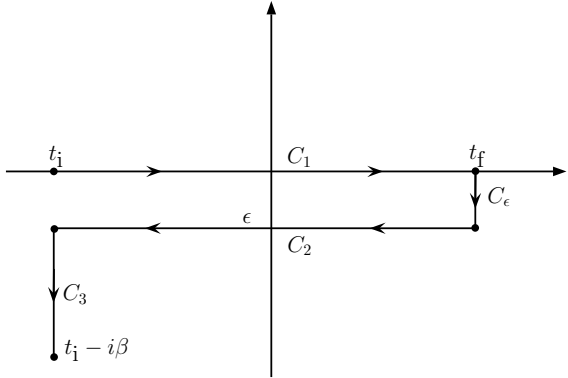}
   \caption{The Schwinger-Keldysh contour $C$ in the complex-time plane. The part $C_1$ is on the real axis and runs forwards in time, while
            the part $C_2$ has a small imaginary part $-i\epsilon$ and runs parallel to the real axis but backwards in time. The part $C_3$
            runs parallel to the imaginary axis, from $t_\textrm{i} - i\epsilon$ down to $t_\textrm{i} - i\beta$. The time $t_\textrm{i}$ is the
            initial time and the final time $t_\textrm{f}$ is assumed to be larger than the time of any external point on $C$ but is
            otherwise arbitrary. In our computations we will always let $\epsilon \to 0^+$, so the path $C_\epsilon$ gives a vanishing
            contribution.}
   \label{fig:schwinger_keldysh_path}
 \end{figure}
\end{center}

Equation~(\ref{eq:correlator_interaction_pic}) can be expressed more concisely by considering the time as a complex variable defined along the Schwinger-Keldysh contour $C$ of Fig.~\ref{fig:schwinger_keldysh_path}. This contour goes from $t_\textrm{i}$ to $t_\textrm{f}$ on the real axis along $C_1$, then from $t_\textrm{f}$ to $t_\textrm{f} - i\epsilon$ along $C_\epsilon$, then back from $t_\textrm{f} - i\epsilon$ to $t_\textrm{i} - i\epsilon$ along $C_2$, and then finally goes from $t_\textrm{i} - i\epsilon$ to $t_\textrm{f} - i\beta$ parallel to the imaginary axis along $C_3$. Since we take the limit $\epsilon \to 0^+$ at the end of the computation, the contribution from $C_\epsilon$ vanishes. It is convenient to define the contour-ordering $T_C$ as
\begin{equation}\label{eq:T_C}
 T_C[A(z_1)B(z_2)] \equiv \left\{
 \begin{array}{l}
  A(z_1)B(z_2),\ \textrm{if}\ z_1\ \textrm{is ahead of}\ z_2\eqend{,}\\
  B(z_2)A(z_1),\ \textrm{if}\ z_2\ \textrm{is ahead of}\ z_1\eqend{,}
 \end{array}
\right.
\end{equation}
for $z_1,z_2 \in C$. It is clear that the contour-ordering $T_C$ corresponds to the time-ordering $T$ if $z_1, z_2 \in C_1$ and to the anti-time-ordering $\bar{T}$ if $z_1, z_2 \in C_2$. By using the definition~(\ref{eq:T_C}), we can express Eq.~(\ref{eq:correlator_interaction_pic}) concisely as
\begin{equation}\label{eq:correlator_interaction_pic_C}
 \langle T[A(t_1)B(t_2)]\rangle
 = \frac{\tr\{\rho_0\, T_C[U_\textrm{I}(C) A_\textrm{I}(t_1) B_\textrm{I}(t_2)]\}}{\tr\{\rho_0 U_\textrm{I}(C)\}}\eqend{,}
\end{equation}
where we have defined
\begin{equation}\label{eq:dyson_series_C}
 U_\textrm{I}(C) = T_C\exp\ch{-i\int_CdzH_\textrm{I}(z)}\eqend{.}
\end{equation}
In the case of $N$ observables, Eq.~(\ref{eq:correlator_interaction_pic_C}) is easily seen to generalize to
\begin{splitequation}\label{eq:correlator_interaction_pic_C_N}
  &\langle T[A_1(t_1)\dots A_N(t_N)]\rangle \\
  &\quad = \frac{\tr\{\rho_0\, T_C[U_\textrm{I}(C) A_{1,\, \textrm{I}}(t_1)\dots
  A_{N,\, \textrm{I}}(t_N)]\}}{\tr\{\rho_0 U_\textrm{I}(C)\}}\eqend{.}
\end{splitequation}
We note that the vertical path $C_3$ in Fig.~\ref{fig:schwinger_keldysh_path} is necessary so Eq.~(\ref{eq:correlator_interaction_pic_C_N}) corresponds to a thermal average with respect to the interacting thermal state.

Let us now consider an interacting massive real quantum scalar field theory in a static spacetime with non-derivative interactions. The perturbative expansion of the time-ordered $N$-point functions in a finite temperature state, Eq.~(\ref{eq:correlator_interaction_pic_C_N}), can be expressed in
terms of functional derivatives of the generating functional $\mathcal{Z}$ of the interacting theory. Thus, we define
\begin{equation}
\mathcal{Z}(\beta,J) = \tr\left\{ e^{-\beta H} T_C\left[\exp\left( - i \int_{C\times \Sigma}d^n x
\,\phi(x)J(x)\right)\right]\right\}.
\end{equation}
Here, the function $J(x)$ is a classical source on $C\times \Sigma$, where $C$ is the complex-time contour of Fig.~\ref{fig:schwinger_keldysh_path} and $\Sigma$ is the spatial section of our static spacetime. We note $\mathcal{Z}(\beta,0) = Z(\beta)$. Let us introduce the notation $J_i \equiv J|_{C_i}$. Then, for $x_1, \dots, x_N \in C_1\times \Sigma$, the time-ordered $N$-point function is given by
\begin{splitequation}\label{eq:time_ordered_N_point_function}
 & \langle T[\phi(x_1)\dots \phi(x_N)]\rangle \\
 &\quad = \left.\frac{i^N}{\mathcal{Z}(\beta,0)}\frac{\delta^N}{\delta J_1(x_1)\dots \delta J_1(x_N)} \mathcal{Z}(\beta,J)\right|_{J=0},
\end{splitequation}
where $\phi$ denotes the Heisenberg field operator. The functional $\mathcal{Z}$ is related to $\mathcal{Z}_0$, the generating functional for the free theory, through
\begin{splitequation}
 & \mathcal{Z}(\beta,J) \\
 &\quad = \sum_{k = 0}^\infty\frac{(-i)^k}{k!}\cc{\int_{C\times \Sigma}dvol_y V\pp{i\frac{\delta}{\delta J(y)}}}^k \mathcal{Z}_0(\beta,J),
\end{splitequation}
where $dvol_x$ is the spacetime volume element. The generating functional of the free theory is given by
\begin{equation}\label{eq:free_generating_functional}
 \mathcal{Z}_0(\beta,J) = Z_0(\beta) e^{-\frac{i}{2}\int_{C\times \Sigma}dvol_xdvol_{x'}J(x)G(x,x')J(x')},
\end{equation}
where we have defined
\begin{equation}\label{eq:G}
 G(x,x') \equiv \tr\{\rho_0 T_C[\phi(x)\phi(x')]\},
\end{equation}
with the contour-ordering $T_C$ given in Eq.~(\ref{eq:T_C}). Here the field $\phi(x)$ is the free field.

\subsection{The in-in perturbation theories in Minkowski spacetime and in the Rindler wedge}

In this subsection we describe the in-in perturbation theories in Minkowski spacetime and in the Rindler wedge. The main aim of this paper is to establish the equivalence of these two perturbation theories.

Recall that the generating functional for time-ordered products of the standard in-out formalism in the vacuum state $|\Omega\rangle$ in Minkowski spacetime is
\begin{splitequation} \label{eq:in-out}
 & \mathcal{Z}_\textrm{M}^{\textrm{in-out}}(\textrm{vac},J) \\
 & \quad = \sum_{k = 0}^\infty\frac{(-i)^k}{k!}\cc{\int_{C_1\times \mathbb{R}^{n-1}}dvol_y V\pp{i\frac{\delta}{\delta J(y)}}}^k\\
 & \quad \quad \times \mathcal{Z}_{\textrm{M},0}^{\textrm{in-out}}(\textrm{vac},J)\eqend{,}
\end{splitequation}
where $C_1$ is defined in Fig.~\ref{fig:schwinger_keldysh_path} with $t_\textrm{i} = - \infty$ and $t_\textrm{f}=\infty$. Thus, the integrals in Eq.~(\ref{eq:in-out}) is over the whole of Minkowski spacetime. The generating functional for the free theory is given by
\begin{equation}\label{eq:free_generating_functional_Minkowski}
 \mathcal{Z}^{\textrm{in-out}}_{\textrm{M},0}(\textrm{vac},J)
 = e^{-\frac{i}{2}\int_{C_1\times \mathbb{R}^{n-1}}dvol_xdvol_{x'}J(x)G^{\textrm{F}}(x,x')J(x')}\eqend{,}
\end{equation}
where $G^{\textrm{F}}(x,x')$ is the Feynman propagator for the scalar field. The time-ordered $N$-point function is given by Eq.~(\ref{eq:time_ordered_N_point_function}) with $\mathcal{Z}(\beta,J)$ and $\mathcal{Z}(\beta,0)$ replaced by $\mathcal{Z}^{\textrm{in-out}}_\textrm{M}(\textrm{vac},J)$ and $\mathcal{Z}^{\textrm{in-out}}_{\textrm{M}}(\textrm{vac},0)$, respectively.

The in-in perturbation theory in Minkowski spacetime may be defined by changing the integration range for the internal vertices in Eqs.~(\ref{eq:in-out}) and (\ref{eq:free_generating_functional_Minkowski}) from $C_1 \times \mathbb{R}^{n-1}$ to $(C_1 \cup C_2)\times \mathbb{R}^{n-1}$, where, again, the paths $C_1$ and $C_2$ are shown in Fig.~\ref{fig:schwinger_keldysh_path}, with $t_\textrm{i}=-\infty$ and $t_\textrm{f}=\infty$. Thus, the generating functionals in the in-in formalism in Minkowski spacetime is
\begin{splitequation} \label{eq:in-in}
&  \mathcal{Z}_\textrm{M}^{\textrm{in-in}}(\textrm{vac},J)\\
& = \sum_{k = 0}^\infty\frac{(-i)^k}{k!}\cc{\int_{(C_1\cup C_2)\times \mathbb{R}^{n-1}}
dvol_y V\pp{i\frac{\delta}{\delta J(y)}}}^k \\
& \ \ \ \times \mathcal{Z}_{\textrm{M},0}^{\textrm{in-in}}(\textrm{vac},J)\eqend{,}
\end{splitequation}
where
\begin{splitequation}\label{eq:free_generating_functional_Minkowski2}
 & \mathcal{Z}^{\textrm{in-in}}_{\textrm{M},0}(\textrm{vac},J) \\
 & \quad = e^{-\frac{i}{2}\int_{(C_1\cup C_2)\times \mathbb{R}^{n-1}}dvol_x dvol_{x'}J(x)G(x,x')J(x')}\eqend{.}
\end{splitequation}
Here, the correlator $G(x,x')$ is defined by Eq.~(\ref{eq:G}). Depending on whether the time coordinates $t$ and $t'$ of the points $x$ and $x'$ are on $C_1$ or $C_2$ in Eq.~(\ref{eq:G}) the correlation function is one of the Wightman functions $G^\pm$, the Feynman propagator $G^\textrm{F}$, or the Dyson
(or anti-Feynman) propagator $G^\textrm{D}$. We find from Eqs.~(\ref{eq:T_C}) and~(\ref{eq:G}) that 
\begin{equation}\label{eq:G_two_point_functions_relation}
 G(x,x') =
 \left\{
 \begin{array}{l}
  G^+(x,x'),\ \textrm{if}\ t\in C_2\ \textrm{and}\ t'\in C_1\eqend{,}\\
  G^-(x,x'),\ \textrm{if}\ t\in C_1\ \textrm{and}\ t'\in C_2\eqend{,}\\
  G^\textrm{F}(x,x'),\ \textrm{if}\ t,t'\in C_1\eqend{,}\\
  G^\textrm{D}(x,x'),\ \textrm{if}\ t,t'\in C_2\eqend{.}
 \end{array}
\right.
\end{equation}

As is well known, the in-in formalism gives the correct $N$-point time-ordered product because $\mathcal{Z}^{\textrm{in-in}}_{\textrm{M}}(\textrm{vac},J_1)^c = \mathcal{Z}^{\textrm{in-out}}_{\textrm{M}}(\textrm{vac},J_1)^c$, where the superscript $c$ indicates that only the connected diagrams are included. We present a proof of this fact under the assumption that the one-point function for the Heisenberg field $\phi(x)$ vanishes,
i.e.\ $\langle \Omega|\phi(x)|\Omega\rangle = 0$. We consider the Fourier transform $F(p_1,p_2,\ldots,p_N)$ of the time-ordered $N$-point function:
\begin{eqnarray}
&& F(p_1,p_2,\ldots,p_N)(2\pi)^n \delta^{(n)}(p_1+p_2+\cdots p_N)\nonumber \\
&& = \int dvol_{x_1} dvol_{x_2}\cdots dvol_{x_N}
e^{-i(p_1\cdot x_1 + p_2\cdot x_2+\cdots p_N\cdot x_N)} \nonumber \\
&&\ \ \ \times \langle \Omega|T\left[\phi(x_1)\phi(x_2)\cdots \phi(x_N)\right]|\Omega\rangle\eqend{,}
\end{eqnarray}
where $\delta^{(n)}(p)$ denotes the $n$-dimensional $\delta$-distribution. Note that the correlator connecting a point $x$ with time coordinate 
$t \in C_1$ to a point $x'$ with time coordinate $t' \in C_2$ is the Wightman two-point function given by
\begin{equation}
G^+(x',x) = \int \frac{d^n k}{(2\pi)^{n-1}}\Theta(k^0) \delta( k^2 +m^2)e^{ik\cdot (x'-x)}\eqend{,}
\end{equation}
where $k^0$ is the $0$th component of the relativistic momentum $k$ and $\Theta(x)$ is the Heaviside step function. (From now on, we say, ``the spacetime point $x$ is on $C_i$'' to mean ``the time coordinate $t$ of the spacetime point $x$ is on $C_i$'' for brevity.) Any diagram contributing to
$F(p_1,p_2,\ldots,p_N)$ with $L$ correlators between $C_1$ and $C_2$ takes the form
\begin{eqnarray}
I & = & \int \left\{\prod_{J=1}^L \frac{d^n k_J}{(2\pi)^n} \Theta(k_J^0)\delta(k_J^2 + m^2)\right\}\nonumber \\
&& \times \mathcal{A}_1(p_1,p_2,\ldots,p_N;k_1,k_2,\ldots, k_L)
\mathcal{A}_2(k_1,k_2,\ldots,k_L)\nonumber \\
&& \times (2\pi)^n \delta^{(n)}(k_1+k_2+\cdots+k_L)\eqend{,}
\end{eqnarray}
where we may let $L\geq 2$ because the assumption $\langle\Omega|\phi(x)|\Omega\rangle =0$ implies that the vacuum bubble connected to only one correlator vanishes. The integral $I$ must vanish because for non-zero contribution the $0$th components of the momenta, $k_1^0,k_2^0,\ldots,k_L^0$, must be positive and satisfy $k_1^0+k_2^0+\cdots+k_L^0=0$ at the same time, which is impossible. Thus, we have $\mathcal{Z}^{\textrm{in-in}}_{\textrm{M}}(\textrm{vac},J_1)^c = \mathcal{Z}^{\textrm{in-out}}_{\textrm{M}}(\textrm{vac},J_1)^c$ if we let $t_\textrm{f}=\infty$ (and $t_\textrm{i}=-\infty$) in the in-in formalism.

Now, the time-ordered $N$-point function in the in-in formalism is independent of $t_\textrm{f}$ as long as it is in the future of the time coordinates of all external $N$ points. This well-known fact was shown indirectly in Sec.~\ref{sec:schwinger-keldysh} but can also be proved diagrammatically as follows. A diagram contributing to a time-ordered $N$-point function in the configuration space with $L$ internal vertices takes the form
\begin{eqnarray}
&&  \Delta(x_1,x_2,\ldots,x_N)\nonumber \\
 &&  = \sum_{i_1=1}^2 \sum_{i_2=1}^2 \cdots \sum_{i_L=1}^2
 \int dvol_{y_1}dvol_{y_2}\cdots dvol_{y_L}\nonumber \\
&& \ \ \ \ \ \times \mathcal{C}(x_1,x_2,\ldots,x_N;y_1^{(i_1)},y_2^{(i_2)},\ldots,y_L^{(i_L)})\eqend{,}
\label{eq:Delta}
\end{eqnarray}
where $y_J^{(1)}$ ($y_J^{(2)}$) indicates that the internal vertex point $y_J$ in Minkowski spacetime is on $C_1$ ($C_2$). Now, suppose that at least one internal vertex is in the future of all external points $x_1,x_2,\ldots,x_N$. Let the internal vertex furthest into the future be $y_1$ without loss of generality. Recall that the correlator connecting two points $z$ and $z'$ is $G^{+}(z,z')$ if $z$ is ahead of $z'$ in the contour-ordering. Since the point $y_1^{(1)}$ is ahead of all other points on $C_1$ and the point $y_1^{(2)}$ is behind all other points on $C_2$ by assumption, the correlators in $\mathcal{C}(x_1,x_2,\ldots,x_N;y_1^{(1)},y_2^{(i_2)},\ldots,y_L^{(i_L)})$ and in $\mathcal{C}(x_1,x_2,\ldots,x_N:y_1^{(2)},y_2^{(i_2)},\ldots,y_L^{(i_L)})$ connecting any two given points are the same. The only difference between these two functions is that the vertex factors at $y_1$ have opposite signs. This sign difference comes from the fact that the path $C_2$ runs backward in time whereas the path $C_1$ runs forward. Hence, 
\begin{eqnarray}
&& \mathcal{C}(x_1,x_2,\ldots,x_N;y_1^{(1)},y_2^{(i_2)},\ldots,y_L^{(i_L)}) \nonumber \\
&& = - \mathcal{C}(x_1,x_2,\ldots,x_N:y_1^{(2)},y_2^{(i_2)},\ldots,y_L^{(i_L)})
\end{eqnarray}
and the functions $ \mathcal{C}(x_1,x_2,\ldots,x_N;y_1^{(i_1)},y_2^{(i_2)},\ldots,y_L^{(i_L)})$ cancel pairwise in Eq.~(\ref{eq:Delta}). Thus, there is no contribution from the configuration-space integral if there are internal vertices in the future of all external points. This in turn shows that the upper limit $t_\textrm{f}$ of the time-integration is arbitrary as long as it is larger than the time coordinates of all external points $x_1,x_2,\ldots,x_N$. Thus, the in-in perturbation theory correctly gives the time-ordered $N$-point functions for any $t_\textrm{f}$ as long as it is in the future of all external points. (Note that diagrams with internal vertices on $C_2$ do contribute if $t_\textrm{f}$ is finite in the in-in formalism.)

Now, the region of Minkowski spacetime with $|x^0| < x^{n-1}$ is called the Rindler wedge. By introducing the coordinates $\tau$ and $\xi$ according to~\cite{rindler_ajp_1966}
\begin{equations}[eq:Rindler_coordinates]
 x^0 & = \frac{e^{a\xi}}{a} \sinh a\tau\eqend{,} \\
 x^{n - 1} & = \frac{e^{a\xi}}{a} \cosh a\tau\eqend{,}
\end{equations}
the metric of the spacetime in the Rindler wedge is given by
\begin{equation}
ds^2 = e^{2a\xi} (- d\tau^2 + d\xi^2) + (dx^1)^2 + \cdots + (dx^{n-2})^2\eqend{.}
\end{equation}
This spacetime is static with $\tau$ as time, and a conserved energy can be defined with respect to the symmetry $\tau \to \tau + \textrm{constant}$, which is a boost symmetry of Minkowski spacetime. This energy is called the Rindler energy. The results of Bisognano and Wichmann~\cite{bisognano_wichmann_jmp_1975} and Unruh and Weiss~\cite{unruh_weiss_prd_1984} imply that, if the points $x_1,x_2,\ldots x_n$ of the $N$-point function $\langle \Omega|T[\phi(x_1)\phi(x_2)\cdots \phi(x_N)]|\Omega\rangle$ is in the Rindler wedge, then this $N$-point function can be obtained using the Schwinger-Keldysh formalism outlined in Sec.~\ref{sec:schwinger-keldysh} with inverse temperature $\beta = 2\pi/a$ with respect to the Rindler energy. In particular, the free thermal correlator~(\ref{eq:G}) in the Rindler wedge is identical to the Minkowski counterpart~(\ref{eq:G_two_point_functions_relation}) in the vacuum state~\cite{unruh_prd_1976} (see also Ref.~\cite{fulling_prd_1973}).

For the Schwinger-Keldysh perturbation theory in the Rindler wedge with $\beta = 2\pi/a$, the time coordinate is $\tau$ and the Cauchy surface $\Sigma$ is
the surface of constant $\tau$ with coordinates $\xi,x^1,\ldots, x^{n-2}$. Now, suppose that the contribution to the $N$-point function coming from the diagrams with some correlators connecting points on $C_1$ or $C_2$ to points on $C_3$ in Fig.~\ref{fig:schwinger_keldysh_path} vanishes in the limit 
$\tau_\textrm{i}\to -\infty$. This condition is called factorization~\cite{landsman_van_weert_pr_1987} because it will imply that, if the source 
$J(x_1),\cdots,J(x_N)$ are either on $C_1$ or $C_2$, then $\mathcal{Z}(\beta,J) \to \mathcal{Z}^{(3)}(\beta)\mathcal{Z}^{(1,2)}(\beta,J)$ in this limit,
where $\mathcal{Z}^{(3)}(\beta)$ consists of diagrams with all internal vertices on $C_3$ and where for $\mathcal{Z}^{(1,2)}(\beta,J)$ the internal vertices are on $C_1 \cup C_2$. Thus, if the factorization property holds, then the $N$-point function is obtained through Eq.~(\ref{eq:time_ordered_N_point_function}) with 
\begin{splitequation} \label{eq:in-in-Rindler}
&  \mathcal{Z}_\textrm{R}^{\textrm{in-in}}(\beta,J)\\
& = \sum_{k = 0}^\infty\frac{(-i)^k}{k!}\cc{\int_{(C_1\cup C_2)\times \Sigma}
dvol_y V\pp{i\frac{\delta}{\delta J(y)}}}^k \\
& \ \ \ \times \mathcal{Z}_{\textrm{R},0}^{\textrm{in-in}}(\beta,J)\eqend{,}
\end{splitequation}
where $\mathcal{Z}_{\textrm{R},0}^{\textrm{in-in}}(\beta,J) = \mathcal{Z}_{\textrm{M},0}^{\textrm{in-in}} (\textrm{vac},J)$, assuming that $J$ has support in the Rindler wedge. We call the perturbation theory with this generating functional the in-in formalism in Rindler wedge. Note that the equality of
the $N$-point functions computed in these two in-in perturbation theories is equivalent to 
\begin{equation}\label{eq:connected-equality}
\mathcal{Z}_\textrm{R}^\textrm{in-in}(\beta,J)^c = \mathcal{Z}_{\textrm{M}}^\textrm{in-in}(\textrm{vac},J)^c\eqend{.}
\end{equation}

An obvious strategy to prove this equality is to demonstrate the factorization property in the Rindler Schwinger-Keldysh perturbation theory. The factorization property is known to hold in similar situations, e.g.\ for the Schwinger-Keldysh perturbation theory in the static patch of de~Sitter spacetime~\cite{higuchi_marolf_morrison_prd_2011}. However, the infinite volume of $\Sigma$ poses some difficulties in the Rindler case. Fortunately it is possible to show Eq.~(\ref{eq:connected-equality}) or, equivalently, the equality of the $N$-point functions in these two perturbation theories directly by using the light-cone coordinates.

Notice that $(C_1\cup C_2)\times \Sigma$ in Eq.~(\ref{eq:in-in-Rindler}), with $C_1\cup C_2$ being with respect to the Rindler time $\tau$, is simply the restriction of $(C_1 \cup C_2) \times \mathbb{R}^{n-1}$, with $C_1 \cup C_2$ being with respect to the Minkowski time $t$, to the Rindler wedge.\footnote{It is not difficult to see that the in-in formalism is independent of the choice of the time variable to define the paths $C_1$ and $C_2$ as long as it increases monotonically toward the future.} In the next section we analyze the in-in perturbative expansion of the vacuum time-ordered $N$-point functions of the quantum field $\phi$ in Minkowski spacetime, diagram-by-diagram. We prove that the integration range for the internal vertices can be restricted to any Rindler wedge containing all their external points. This will establish the equality~(\ref{eq:connected-equality}) directly.

\section{The equivalence between the Minkowski and Rindler in-in perturbation theories}                                                              %
\label{sec:equivalence_rindler_minkowski}                                                                                                            %

Recall that the in-in and in-out perturbative approaches coincide for the vacuum state in Minkowski spacetime. What we will show is that the usual in-out perturbation theory in Minkowski spacetime is equivalent to the in-in perturbation theory in the Rindler wedge.

We first demonstrate that the integration range over the internal vertices in the in-out formalism can be restricted to the region bounded by two parallel null planes. For this purpose it is convenient to employ the retarded and advanced light-cone coordinates 
\begin{equations}[eq:light_cone_coordinates]
u & \equiv \frac{x^0 - x^{n - 1}}{\sqrt{2}}
\eqend{,} \\
v & \equiv \frac{x^0 + x^{n - 1}}{\sqrt{2}},
\end{equations}
respectively. We also let $\vt{v}_\bot$ denote the part of a spatial vector $\vt{v}$ transverse to the $x^{n - 1}$-direction. In terms of these
coordinates, the free Feynman propagator reads 
\begin{splitequation}\label{eq:Feynman_propagator_lightcone}
 G^\textrm{F}(x,x') & = \frac{1}{(2\pi)^{n - 1}}\int_{\mathbb{R}^{n - 1}}d\kappa d^{n - 2}\mathbf{k}_\bot \tilde{G}^\textrm{F}(v - v',\kappa,\vt{k}_\bot)\\
 &\qquad \qquad \qquad \qquad \times e^{i[\vt{k}_\bot \cdot (\vt{x}_\bot - \vt{x}_\bot') - \kappa(u - u')]}\eqend{,}
\end{splitequation}
where
\begin{splitequation}\label{eq:G_F_tilde}
 \tilde{G}^\textrm{F}(v,\kappa,\vt{k}_\bot) & \equiv \frac{1}{2|\kappa|}[\Theta(v)\Theta(\kappa) + \Theta(-v)\Theta(-\kappa)] \\
&\ \times \exp\ch{-i \frac{\vt{k}_\bot^2 + m^2}{2\kappa}v}\eqend{.}
\end{splitequation}
Here, $\kappa$ is the momentum conjugate to the light-cone coordinate $u$, which we call $u$-energy in this paper, and $\vt{k}_\bot$ is the transverse momentum. The form of $\tilde{G}^\textrm{F}$ can be obtained either as the Fourier transform of $G^\textrm{F}$ with respect to $u$ and $\vt{x}_\bot$, written in terms of the usual momentum space expression, or by quantizing the free theory in the light-cone coordinates $u,v$. We briefly discuss the second approach in Appendix~\ref{sec:appendix_light_cone_quantisation}.

We start with discussing some general features of the perturbative expansion of the $N$-point function $\langle \Omega|T[\phi(x_1)\phi(x_2)\cdots\phi(x_N)]|\Omega\rangle$ in light-cone coordinates. Let $x_i = (v_i^{(e)}, u_i^{(e)}, \mathbf{x}^{(e)}_{i\perp})$ be the external point coordinates. We consider the Fourier transform of the $N$-point function with respect to the $u$- and $\mathbf{x}_\perp$-coordinates as 
\begin{eqnarray}
&& A(v_1^{(e)},\ldots,v_N^{(e)};\kappa_1^{(e)},\ldots,\kappa_N^{(e)};\mathbf{p}_{1\perp}^{(e)},\ldots,\mathbf{p}_{N\perp}^{(e)}) \nonumber \\
&& \times
(2\pi)^{n-1}\delta\left(\sum_{i=1}^N \kappa_i^{(e)}\right)\delta^{(n-2)}\left(\sum_{i=1}^N \textbf{p}_{i\perp}^{(e)}\right)
 \nonumber \\
&& = \int \prod_{j=1}^N du_i^{(e)} d^{n - 2}\vt{x}_{i\perp}^{(e)}
\exp\left[ i\sum_{i=1}^N \left( \kappa_i^{(e)} u_i^{(e)} - \mathbf{p}_{i\perp}^{(e)}\cdot \mathbf{x}_{i\perp}^{(e)}\right)\right]
\nonumber \\
&& \ \ \times \langle \Omega|T\left[\phi(x_1)\cdots\phi(x_N)\right]|\Omega\rangle\eqend{.}
\label{eq:amplitude_definition}
\end{eqnarray}
The diagrammatic expansion of the amplitude
$A\equiv A(v_1^{(e)},\ldots,v_N^{(e)};\kappa_1^{(e)},\ldots,\kappa_N^{(e)};\mathbf{p}_{1\perp}^{(e)},\ldots,\mathbf{p}_{N\perp}^{(e)})$ is analogous
to that in the conventional approach. We note that 
the small-$\kappa$ 
behavior of the internal propagators is milder than one might expect from 
the factor $1/|\kappa|$ in Eq.~(\ref{eq:G_F_tilde}) if the $v$-coordinates of the internal points are distinct because of the oscillatory factor. This can be seen by changing the variable as $y=1/\kappa$, which results in a decreasing and oscillatory integrand for large $y$ (i.e.\ for small $\kappa$). We also note that one may let all $v$-coordinates of the vertices be distinct before integrating over these coordinates because there is no delta-function-like contribution at $v=0$ in Eq.~(\ref{eq:G_F_tilde}).

Each vertex, internal or external, in a diagram contributing to the amplitude $A$ in Eq.~(\ref{eq:amplitude_definition}) carries a $v$-coordinate and is multiplied by a suitable factor representing the interaction there. (The integral over the $v$-coordinates of the internal vertices will be carried out
in the end.) The vertices are connected by the propagator $\tilde{G}^\textrm{F}(v,\kappa,\mathbf{k}_\perp)$, where $v$ is the difference between the $v$-coordinates connected by this propagator. Equation (\ref{eq:G_F_tilde}) shows that it does not matter whether $v$ is chosen to be positive or negative.  We choose $v>0$. Then the $u$-energy $\kappa$ must be positive for $\tilde{G}^\textrm{F}(v,\kappa,\mathbf{k}_\perp)$ to be non-zero. Thus, $u$-energy can be regarded to flow toward the vertex with the higher value of $v$. Like the usual energy, the $u$-energy is conserved at each internal vertex.

Now, suppose all external points $x_1, x_2,\ldots, x_N$ are in the spacetime region between the two null planes $v=V_1$ and $v=V_2$ with $V_1 < V_2$, which is denoted by $O(V_1,V_2)$. We will show that the integration over the internal vertices can be restricted to the region $O(V_1,V_2)$, assuming that the vacuum expectation value $\langle \Omega|\phi(x)|\Omega\rangle$ vanishes. Suppose that one or more internal vertices of a diagram contributing to $A$ have the $v$-coordinates larger than or equal to $V_2$. Since $\langle \Omega|\phi(x)|\Omega\rangle = 0$, the sum of all tadpole subdiagrams vanishes. Hence we may assume that two or more propagators connect points in the future of or on the null plane $v=V_2$ to points in its past. Let these propagators be
$\tilde{G}^\textrm{F}(\tilde{v}_i - \check{v}_i,\kappa_i,\mathbf{k}_{i\perp})$, $i=1,2,\ldots,L$, where $\tilde{v}_i \geq V_2 > \check{v}_i$. Then, from Eq.~(\ref{eq:G_F_tilde}) we have
\begin{splitequation}
& \tilde{G}^\textrm{F}(\tilde{v}_i-\check{v}_i,\kappa_i,\mathbf{k}_{i\perp})\\
& \qquad = \frac{\Theta(\kappa_i)}{2\kappa_i}\exp\left\{ - i \frac{\mathbf{k}^2_{i\perp}+m^2}{2\kappa_i}
(\tilde{v}_i - \check{v}_i)\right\}\eqend{.}
\end{splitequation}
Thus, the contribution of this diagram to the amplitude $A$ before the integration over the $v$-coordinates of the internal vertices and over the transverse momenta takes the form
\begin{eqnarray}
\tilde{A} & = & \prod_{i=1}^L \int_0^\infty \frac{d\kappa_i}{2\kappa_i}
\exp\left\{ -i \frac{\mathbf{k}_{i\perp}^2 + m^2}{2\kappa_i}(\tilde{v}_i - \check{v}_i)
\right\} \nonumber \\
&& \times
F_1(\tilde{v}_1,\ldots,\tilde{v}_L; \kappa_1,\ldots,\kappa_L;\mathbf{k}_{1\perp},\ldots,\mathbf{k}_{L\perp})
\nonumber\\
&& \times
F_2(\check{v}_1,\ldots,\check{v}_L;\kappa_1,\ldots,\kappa_L;\mathbf{k}_{1\perp},\ldots,\mathbf{k}_{L\perp})
\nonumber \\
&& \times \delta(\kappa_1+\kappa_2+\cdots+\kappa_L)\eqend{,}
\end{eqnarray}
where $F_2(\check{v}_1,\ldots,\check{v}_L;\kappa_1,\ldots,\kappa_L;\mathbf{k}_{1\perp},\ldots,\mathbf{k}_{L\perp})$ depends also on $v_i^{(e)}$,
$\kappa_i^{(e)}$ and $\mathbf{p}_{i\perp}^{(e)}$, $i=1,2,\ldots,N$. The quantity $\tilde{A}$ must vanish because $\kappa_1$, $\kappa_2$, $\ldots$ and 
$\kappa_L$, which are positive, must add up to zero for a non-zero contribution to $\tilde{A}$.

One can show similarly that the contribution to the amplitude $A$ from any diagram with one or more points in the past of or on the null plane $v=V_1$ vanishes. 
Thus, the amplitude $A$ can be calculated with the integration region for the internal vertices restricted to $O(V_1,V_2)$.

So far we have shown that the vertex integration can be restricted to the region $O(V_1,V_2)$ for the in-out Feynman diagrams, i.e.\ for the diagrams such that all internal vertices (as well as the external points) are on $C_1$ of the contour $C$ in Fig.~\ref{fig:schwinger_keldysh_path}. We now use this result to show that the integration for the internal vertices can be restricted to any Rindler wedge containing all external points in the in-in perturbation theory.

First we note that the internal vertices for the in-in perturbation theory can also be restricted to the region $O(V_1,V_2)$ because the contribution to $A$ of any diagram in the in-in perturbation theory with one or more points on $C_2$ in Fig.~\ref{fig:schwinger_keldysh_path}, where $C_1$ and $C_2$ are complex paths in the variable $u$ in light-cone coordinates, vanishes. This can be proved by the same argument as that for proving the equivalence of the in-in and in-out perturbation theories in the full Minkowski spacetime with the energy conservation replaced by the $u$-energy conservation. Thus, the time-ordered $N$-point function can be calculated with the internal vertices restricted to the region between two null planes which contains all external points in the in-in formalism as well.

Now, by spacetime translation invariance of Minkowski spacetime we may assume that all external points are contained in the Rindler wedge $|x^0| < x^{n-1}$ without loss of generality. Then, the time-ordered $N$-point function can be calculated using the in-in perturbation theory with the internal vertices restricted to the region between the two null planes $v=V_2 > 0$ and $v=0$ that contains all external points. Since $V_2$ is arbitrary as long as it is larger than the $v$-coordinate of any external point, we can let the integration region for the internal vertices be the half space $v> 0$. Then, by the same argument as for the proof that the $t$-integral can be restricted by $t < t_\textrm{f}$ for any $t_\textrm{f}$ in the future of all external points in the in-in perturbation theory in Minkowski spacetime, it can be shown that one can restrict the $u$-integration to the region $u < u_\textrm{f}$, where $u_\textrm{f}$ is larger than the $u$-coordinate of any external point but otherwise arbitrary. In particular we can require $u<0$. Thus, the time-ordered $N$-point function can be calculated with the internal vertices restricted to the region satisfying $v = (x^0 + x^{n-1})/\sqrt{2} >0$ and $(x^0 - x^{n-1})/\sqrt{2} < 0$, i.e.\ the Rindler wedge satisfying $|x^0| < x^{n-1}$. In other words, the in-in perturbation theory in Minkowski spacetime for the vacuum state is equivalent to that in the Rindler wedge in the thermal state with Unruh temperature $a/2\pi$.

\section{Example for the restriction of integration over the internal vertices}                                                                      %
\label{sec:explicit_example}                                                                                                                         %

\begin{center}
 \begin{figure}[ht]
   \includegraphics[scale=0.85]{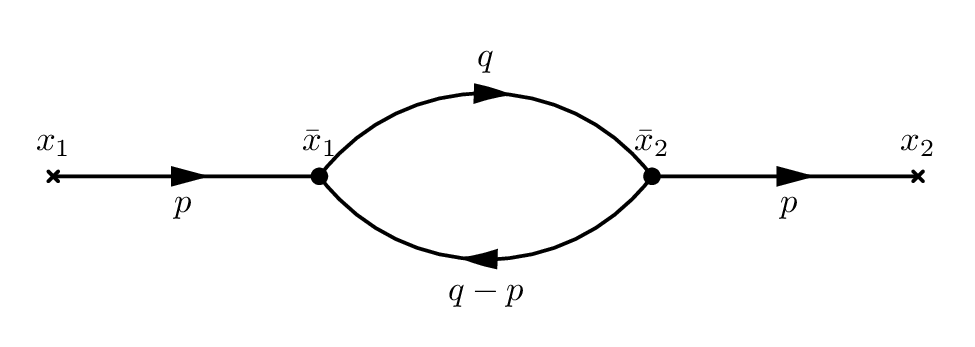}
   \caption{One-loop correction to the propagator in a $\phi^3$ theory. The momentum $q$ circulates in the loop, while $p$ denotes the external 
            momentum. The points $\bar{x}_1$ and $\bar{x}_2$ denote the internal vertices and $x_1$ and $x_2$ the external points.}
   \label{fig:one_loop_feynman_diagram}
 \end{figure}
\end{center}

Our proof of the equivalence between the two in-in perturbation theories relied on the result that the integration region for the internal vertices can be
restricted to the region between two null planes that contains all external points in the in-out perturbation theory. In this section we demonstrate this result for the diagram shown in Fig.~\ref{fig:one_loop_feynman_diagram} in the $\phi^3$-theory. (We omit the factor $i\lambda$, if the interaction term in the Lagrangian density is $i\lambda\phi^3$, at the vertices.)

We first consider this diagram in the whole of Minkowski spacetime. If the external momentum is $(p^0,\mathbf{p}_\perp,p^{n-1})$, then the truncated two-point function at one-loop reads
\begin{equation}
A(p,\mathbf{p}_\perp) = \int \frac{d^{n-2}\mathbf{q}_\perp}{(2\pi)^{n-2}}\, \Pi(p)\eqend{,}
\end{equation}
where
\begin{equation}\label{eq:Pi-p-1}
 \Pi(p) = - \int_{\mathbb{R}^2}\frac{d^2q}{(2\pi)^2}
\frac{1}{[q^2 + m_1^2-i\epsilon][(p - q)^2 + m_2^2-i\epsilon]}\eqend{.}
\end{equation}
(The truncated two-point function $\Pi(p)$ includes the factor $i^2$ from the propagators.) Here, we have defined $p = (p^0,p^{n-1})$, $q=(q^0,q^{n-1})$,
$m_1^2 = m^2 + \mathbf{q}_\perp^2$ and $m_2^2 = m^2 + (\mathbf{p}_\perp - \mathbf{q}_\perp)^2$. We also have defined $q^2 = -(q^0)^2 + (q^{n-1})^2$ and
$(p-q)^2 = -(p^0 - q^0)^2 + (p^{n-1} - q^{n-1})^2$. After the standard Feynman parametrization, one can readily perform the $q$-integral after the Wick rotation $q^0 \to iq^0$. The result is
\begin{equation}\label{eq:Pi_Lorentzian}
 \Pi(p) = \frac{i}{4\pi}\int_0^1 \frac{d\alpha}{\alpha(1 - \alpha)} \frac{1}{2p^+p^- - m^2(\alpha) + i\epsilon}\eqend{.}
\end{equation}
where we have defined $p^\pm \equiv (p^0 \pm p^{n-1})/\sqrt{2}$ and
\begin{equation}\label{eq:m_squared}
 m^2(\alpha) \equiv \frac{(1 - \alpha)m_1^2 + \alpha m^2_2}{\alpha(1 - \alpha)}\eqend{.}
\end{equation}

Now, we attach the propagators to the truncated two-point function and then perform the inverse Fourier transform to convert the variables $p^\pm$ to the configuration variables $(u,v)$. (We do not perform the inverse Fourier transform for the transverse momentum $\mathbf{p}_\perp$.)
More precisely, we multiply $\Pi(p)$ in Eq.~(\ref{eq:Pi_Lorentzian}) by $e^{-ip^+(u_1-u_2)-ip^-(v_1 - v_2)}/(2\pi)^2$ and integrate over $p^+$ and $p^-$.  Thus, we have
\begin{splitequation}\label{eq:D_M_expression}
 D_\textrm{M}&(u_1 - u_2,v_1 - v_2) \\
 & = \int_{-\infty}^\infty \frac{dp^+}{2\pi} F_\textrm{M}(v_1 - v_2,p^+)e^{-ip^+(u_1 - u_2)}\eqend{,}
\end{splitequation}
where
\begin{eqnarray}
 F_\textrm{M}(v,p^+) &\equiv & -\frac{i}{4\pi}\int_0^1 \frac{d\alpha}{\alpha(1 - \alpha)}\int_{-\infty}^\infty
\frac{dp^-}{2\pi} e^{-ip^-v} \nonumber \\
 &&\quad \times \frac{1}{2p^+p^- - M_1^2 + i\epsilon} \frac{1}{2p^+p^- - m^2(\alpha) + i\epsilon}
\nonumber \\
 && \quad \times \frac{1}{2p^+p^- - M_2^2 + i\epsilon}\eqend{.} \label{eq:F_definition}
\end{eqnarray}
Here, we have defined $M_1^2 = M_2^2 \equiv m^2 + \mathbf{p}_\perp^2$. [The minus sign in Eq.~(\ref{eq:F_definition}) comes from the factor of $i^2$ from the two propagators attached.] We find it convenient to let $M_1 \neq M_2$. The equality of the two perturbation theories we are comparing turns out to hold even if $M_1 \neq M_2$ and remain to hold in the limit $M_2 \to M_1$. What we will show is that the function $F_\textrm{M}(v_1-v_2,p^+)$ is reproduced with the internal vertices restricted to the region $O(V_1,V_2)$ between the two null planes $v=V_1$ and $v=V_2$ if $v_1, v_2 \in (V_1,V_2)$.

We perform the $p^{-}$-integral in Eq.~(\ref{eq:F_definition}) using the residue theorem. All three poles lie in the lower half-plane if $p^{+}>0$ and in the upper half-plane if $p^{+} < 0$. The contour is closed in the lower half-plane if $v>0$ and in the upper half-plane if $v < 0$ because of the factor $e^{-ip^{-}v}$. Thus, we find
\begin{splitequation}\label{eq:F_integrated}
 F_\textrm{M} &= -\frac{1}{8\pi|p^+|}[\Theta(v)\Theta(p^+) + \Theta(-v)\Theta(-p^+)]\\
 &\quad \times \int_0^1 \frac{d\alpha}{\alpha(1 - \alpha)}\left\{\frac{e^{-i\frac{M_1^2}{2p^+}v}}{[M_1^2 - m^2(\alpha)](M_1^2 - M_2^2)} \right. \\
 &\qquad + \frac{e^{-i\frac{m^2(\alpha)}{2p^+}v}}{[m^2(\alpha) - M_1^2][m^2(\alpha) - M_2^2]} \\
 &\qquad \left. + \frac{e^{-i\frac{M_2^2}{2p^+}v}}{(M_2^2 - M_1^2)[M_2^2 - m^2(\alpha)]} \right\}.
\end{splitequation}
Note that the limit $M_1^2 - M_2^2 \to 0$ in Eq.~(\ref{eq:F_integrated}) is non-singular as the singularities in the first and third terms cancel out in this limit. Similarly, the limits $M_1^2 - m^2(\alpha) \to 0$ and $M_2^2 - m^2(\alpha) \to 0$ also give non-singular results. Also, there is no divergence in the $\alpha$-integration at $\alpha=0$ or $1$ because $\alpha(1-\alpha)m^2(\alpha)$ is non-zero for $\alpha=0$ or $1$.

We now compute the function corresponding to $D_\textrm{M}(u_1-u_2,v_1-v_2)$ with the internal vertices restricted to the region between the null planes $v=V_1$ and $v=V_2$ with $v_1,v_2 \in (V_1,V_2)$. The two-point function to be compared with $D_\textrm{M}(u_1-u_2,v_1-v_2)$ is $D_\textrm{R}(u_1-u_2,v_1-v_2)$ defined by
\begin{eqnarray}
&& D_\textrm{R}(u_1-u_2,v_1-v_2) \nonumber \\
 && = \int_{V_1< \bar{v}_1,\bar{v}_2 < V_2}
d^2\bar{x}_1d^2\bar{x}_2 G^\textrm{F}_{M_1}(x_1, \bar{x}_1) G^\textrm{F}_{m_1}(\bar{x}_1, \bar{x}_2) \nonumber \\
 &&\quad \times G^\textrm{F}_{m_2}(\bar{x}_1, \bar{x}_2) G^\textrm{F}_{M_2}(\bar{x}_2, x_2)\eqend{,}
\end{eqnarray}
where $G^F_m(x,y)$ is the time-ordered two-point function for a scalar field with mass $m$ in two dimensions. Here, $x_i = (u_i,v_i)$ and 
$\bar{x}_i=(\bar{u}_i,\bar{v}_i)$, $i=1,2$. We substitute the expression (\ref{eq:Feynman_propagator_lightcone}) for the Feynman propagator with $n=2$ and find
\begin{widetext}
\begin{splitequation}
 D_\textrm{R} &= \frac{1}{(2\pi)^4}\int_{V_1}^{V_2}d\bar{v}_1 d\bar{v}_2 \int_{-\infty}^\infty d\bar{u}_1d\bar{u}_2 d\kappa_1 d\kappa_2 d\bar{\kappa}_1 d\bar{\kappa}_2
\frac{1}{16|\kappa_1 \bar{\kappa}_1 \bar{\kappa}_2 \kappa_2|}[\Theta(v_1 - \bar{v}_1)\Theta(\kappa_1) + \Theta(\bar{v}_1 - v_1)\Theta(-\kappa_1)]\\
& \quad \times [\Theta(\bar{v}_1 - \bar{v}_2)\Theta(\bar{\kappa}_1)\Theta(\bar{\kappa}_2) + \Theta(\bar{v}_2 - \bar{v}_1)\Theta(-\bar{\kappa}_1)\Theta(-\bar{\kappa}_2)] [\Theta(\bar{v}_2 - v_2)\Theta(\kappa_2) + \Theta(v_2 - \bar{v}_2)\Theta(-\kappa_2)]\\
& \quad \times \exp\left\{-i\frac{M_1}{2\kappa_1}(v_1 - \bar{v}_1) - i\left(\frac{m_1}{2\bar{\kappa}_1} + \frac{m_2}{2\bar{\kappa}_2}\right)(\bar{v}_1 - \bar{v}_2) - i\frac{M_2}{2\kappa_2}(\bar{v}_2 - v_2)\right\} e^{-i[\kappa_1(u_1 - \bar{u}_1) + (\bar{\kappa}_1 + \bar{\kappa}_2)(\bar{u}_1 - \bar{u}_2) + \kappa_2(\bar{v}_2 - v_2)]}\eqend{.}
\end{splitequation}
The integration over the $u$-coordinates $\bar{u}_1$ and $\bar{u}_2$ of the internal vertices produces the factor 
$(2\pi)^2\delta(\bar{\kappa}_1 + \bar{\kappa}_2 - \kappa_1)\delta(\kappa_2 - \bar{\kappa}_1 - \bar{\kappa}_2)$. This makes the integration over 
$\kappa_2$ and $\bar{\kappa}_2$ trivial. Thus, we obtain 
\begin{splitequation}\label{eq:D_R_expression}
 D_\textrm{R}&(u_1 - u_2, v_1 - v_2) = \int_{-\infty}^\infty\frac{d\kappa_1}{2\pi}F_\textrm{R}(v_1 - v_2,\kappa_1)e^{-i\kappa_1(u_1 - u_2)}\eqend{,}
\end{splitequation}
where
 \begin{splitequation}\label{eq:new_F}
 F_\textrm{R}(v_1 - v_2, \kappa_1) &\equiv \int_{V_1}^{V_2}d\bar{v}_1d\bar{v}_2\int_{-\infty}^\infty\frac{d\bar{\kappa}_1}{2\pi}\frac{1}{16\kappa_1^2|\bar{\kappa}_1(\kappa_1 - \bar{\kappa}_1)|} [\Theta(v_1 - \bar{v}_1)\Theta(\bar{v}_1 - \bar{v}_2)\Theta(\bar{v}_2 - v_2)\Theta(\kappa_1)\Theta(\bar{\kappa}_1)\Theta(\kappa_1 - \bar{\kappa}_1) \\
 & \quad + \Theta(\bar{v}_1 - v_1)\Theta(\bar{v}_2 - \bar{v}_1)\Theta(v_2 - \bar{v}_2)\Theta(-\kappa_1)\Theta(-\bar{\kappa}_1)\Theta(\bar{\kappa}_1 - \kappa_1)] \\
 & \quad \times \exp\left\{-i\frac{M_1^2}{2\kappa_1}v_1 - i\frac{M_2^2}{2\kappa_1}v_2 - i\left(\frac{m_1^2}{2\bar{\kappa}_1} + \frac{m^2_2}{2(\kappa_1 - \bar{\kappa}_1)} - \frac{M_1^2}{2\kappa}\right)\bar{v}_1 - i\left(\frac{M_2^2}{2\kappa_1} - \frac{m_1^2}{2\bar{\kappa}_1} - \frac{m^2_2}{2(\kappa_1 - \bar{\kappa}_1)}\right)\bar{v}_2\right\}\eqend{.}
\end{splitequation}
\end{widetext}
Comparing Eqs.~(\ref{eq:D_M_expression}) and (\ref{eq:D_R_expression}), we see that our task is to show that 
$F_\textrm{M}(v_1-v_2;p^+) = F_\textrm{R}(v_1-v_2,p^+)$.

The first term within the brackets in Eq.~(\ref{eq:new_F}) is non-zero only if $v_1 > \bar{v}_1 > \bar{v_2} > v_2$, while the second term gives a non-vanishing contribution only if $v_2 > \bar{v}_2 > \bar{v}_1 > v_1$. Hence, Eq.~(\ref{eq:new_F}) can written as
\begin{splitequation}
 F_\textrm{R}(v_1 - v_2, \kappa_1) &= \Theta(v_1 - v_2)\Theta(\kappa_1)K_1(v_1,v_2,\kappa_1) \\
 & \quad +\Theta(v_2 - v_1)\Theta(-\kappa_1)K_2(v_1,v_2,\kappa_1),
\end{splitequation}
where the functions $K_1(v_1,v_2,\kappa_1)$ and $K_2(v_1,v_2,\kappa_1)$ are defined by
\begin{splitequation}\label{eq:K_1}
 K_1(v_1,v_2,\kappa_1) &\equiv \frac{1}{32\pi\kappa_1^3}\int_{v_2}^{v_1}d\bar{v}_1\int_{v_2}^{\bar{v}_1}d\bar{v}_2\int_0^1\frac{d\alpha}{\alpha(1 - \alpha)}\\
 & \quad \times \exp\left\{-i\frac{M_1^2}{2\kappa_1}v_1 - i\frac{M_2^2}{2\kappa_1}v_2\right. \\
 & \quad \left. - i\frac{m^2(\alpha) - M_1^2}{2\kappa_1}\bar{v}_1 - i\frac{M_2^2 - m^2(\alpha)}{2\kappa_1}\bar{v}_2\right\}\eqend{,}
\end{splitequation}
and
\begin{splitequation}\label{eq:K_2}
 K_2(v_1,v_2,\kappa_1) &\equiv \frac{1}{32\pi|\kappa_1|^3}\int_{v_1}^{v_2}d\bar{v}_2\int_{v_1}^{\bar{v}_2}d\bar{v}_1\int_0^1\frac{d\alpha}{\alpha(1 - \alpha)}\\
 & \quad \times \exp\left\{-i\frac{M_1^2}{2\kappa_1}v_1 - i\frac{M_2^2}{2\kappa_1}v_2\right. \\
 & \quad \left. - i\frac{m^2(\alpha) - M_1^2}{2\kappa_1}\bar{v}_1 - i\frac{M_2^2 - m^2(\alpha)}{2\kappa_1}\bar{v}_2\right\}\eqend{.}
\end{splitequation}
Here, we have changed the integration variable $\bar{\kappa}_1$ to $\alpha$ by letting $\bar{\kappa}_1 = \alpha \kappa_1$. The function $m^2(\alpha)$ was defined by Eq.~(\ref{eq:m_squared}).

The integrals over $\bar{v}_1$ and $\bar{v}_2$ can readily be evaluated using 
\begin{splitequation}
 &\int_{v_2}^{v_1}d\bar{v}_1\int_{v_2}^{\bar{v}_1}d\bar{v}_2 e^{-iB_1\bar{v}_1 - iB_2\bar{v}_2} \\
 & = -\frac{e^{-i(B_1 + B_2)v_1}}{B_2(B_1 + B_2)} + \frac{e^{-i(B_1v_1 + B_2v_2)}}{B_1B_2} - \frac{e^{-
i(B_1 + B_2)v_2}}{B_1(B_1 + B_2)}\eqend{,}
\end{splitequation}
and we indeed find that $F_\textrm{R}(v_1-v_2,p^+)=F_\textrm{M}(v_1-v_2,p^+)$ where $F_\textrm{M}(v_1-v_2,p^+)$ is given by Eq.~(\ref{eq:F_integrated}).

\section{Discussion}                                                                                                                                 %
\label{sec:discussion}                                                                                                                               %

In this paper we showed that the time-ordered $N$-point function for self-interacting massive scalar field in the Minkowski vacuum state can be computed in the in-in formalism with the internal vertices restricted to a Rindler wedge containing all $N$ external points to all orders in perturbation theory. This means that this $N$-point function can be computed as that in the thermal state (with respect to the Rindler time) at the Unruh temperature in the Rindler wedge using the in-in formalism, which is defined as the Schwinger-Keldysh formalism with factorization property, i.e.\ with no contribution from the path $C_3$ in Fig.~\ref{fig:schwinger_keldysh_path}. It is well known that this Schwinger-Keldysh perturbation theory (without the assumption of factorization) reproduces this $N$-point function. Thus, our result is an indirect proof of the factorization property of the Schwinger-Keldysh perturbation theory in the thermal state in the Rindler wedge.

Of course, the amplitude we studied will be plagued with ultraviolet divergences after the vertex integrations are performed in general, and requires some regularization scheme followed by the renormalization of the bare action defining the model. Since our results do not depend on the dimension of Minkowski spacetime, they remain valid if the dimensional regularization is employed. Alternatively, we could make use of the Pauli-Villars regularization method, in which case some regulator fields with large masses are introduced. Clearly, this regularization method will not alter the conclusions of 
Sec.~\ref{sec:equivalence_rindler_minkowski}. Hence, our results there will also hold for the fully renormalized theory.

Our proof of the equivalence between the Minkowski and Rindler in-in perturbation theories for the Minkowski vacuum was presented only for (interacting) scalar field theory. Nevertheless, the proof did not use properties specific to scalar field theory, and we expect that our result will hold for other interacting models involving higher-spin fields. If this is the case, the computation of rates of the same particle physics process in the inertial and
accelerated frames, for example, cannot differ (if the in-in formalism is used), contrary to recent claims~\cite{ahluwalia_labun_torrieri_ejpa_2016} in the context of mixing neutrinos (see Ref.~\cite{cozzella_et_al_prd_2018} for a discussion on this particular issue).

\acknowledgments

W.~L.~acknowledges financial support from Coordena{\c c}{\~a}o de Aperfei{\c c}oamento de Pessoal de N{\'\i}vel Superior (CAPES) through the Graduate Programme in Physics of the Universidade Federal do ABC. This work was supported in part by the Grant  No.~RPG-2018-400, ``Euclidean and in-in formalisms in static spacetimes with Killing horizons'', from the Leverhulme Trust.

\appendix

\section{Quantization of the Klein-Gordon field in the light-cone coordinates}                                                                        %
\label{sec:appendix_light_cone_quantisation}                                                                                                         %

In this appendix we present the quantization of the Klein-Gordon field in terms of the light-cone coordinates $u$ and $v$. This quantization method
dates back to Dirac~\cite{dirac_rmp_1949} and goes under different names in the literature, such as light-cone quantization or light-front
quantization, and is sometimes employed in the analysis of bound states in nuclear physics~\cite{brodsky_pauli_pinsky_pr_1998, heinzl_lnp_2001}.

In the inertial coordinates, the free Klein-Gordon field equation
\begin{equation}\label{eq:KG_eq}
 (-\del_\mu\del^\mu + m^2)\phi = 0\eqend{,}
\end{equation}
admits the normalized positive-frequency modes
\begin{equation}\label{eq:inertial_positive_frequency_modes}
 f_\vt{k}(x) \equiv \frac{1}{(2\pi)^\frac{n - 1}{2}}\frac{e^{i(\vt{k} \cdot \vt{x} - \omega_\vt{k}x^0)}}{\sqrt{2\omega_\vt{k}}}\eqend{,}
\end{equation}
with $\omega_\vt{k} \equiv \sqrt{\vt{k}^2 + m^2}$. In term of the light-cone coordinates~(\ref{eq:light_cone_coordinates}), the
modes~(\ref{eq:inertial_positive_frequency_modes}) read
\begin{equation}\label{eq:inertial_positive_frequency_modes_uv}
 f_\vt{k}(x) \equiv \frac{1}{(2\pi)^\frac{n - 1}{2}}\frac{e^{i\vt{k}_\bot \cdot \vt{x}_\bot}}{\sqrt{2\omega_\vt{k}}}e^{-i(k^+u + k^-v)}\eqend{,}
\end{equation}
where we have defined $k^\pm \equiv (\omega_\vt{k} \pm k^{n - 1})/\sqrt{2}$. Since $\omega_\vt{k} > |k^{n - 1}|$,
the modes~(\ref{eq:inertial_positive_frequency_modes}) are of positive frequency with respect to both $u$ and $v$.

The Klein-Gordon inner product for the solutions to Eq.~(\ref{eq:KG_eq}) is
\begin{equation}\label{eq:KG_norm}
 (f,g)_\textrm{KG} \equiv -i \int_{\mathbb{R}^{n - 1}} d^{n - 1}\vt{x}[g(x)\del_0f^*(x) - \del_0g(x)f^*(x)]\eqend{.}
\end{equation}
The solutions $f_\vt{k}(x)$ are normalized in the sense that
\begin{equation}
(f_\vt{k},f_{\vt{k}'})_\textrm{KG} = \delta^{(n-1)}(\vt{k}-\vt{k}')\eqend{.}
\end{equation}
If we write $k^+=\kappa$, then since $2k^+k^- - \vt{k}_\bot^2 = m^2$, we have
$k^{-} = (\vt{k}_\bot^2 + m^2)/(2\kappa)$. Noting that
\begin{equation}
\delta(\kappa-\kappa')  =  \frac{\omega_{\mathbf{k}}}{\kappa}\delta(k^{n-1} - k^{\prime n-1})\eqend{,}
\end{equation}
the modes proportional to $f_\vt{k}(x)$,
\begin{equation}\label{eq:light_cone_modes_normalized}
 g_{\kappa,\vt{k}_\bot}(x) = \frac{1}{(2\pi)^\frac{n - 1}{2}}\frac{e^{i(\vt{k}_\bot \cdot \vt{x}_\bot - \kappa u)}}{\sqrt{2\kappa}}
 \exp\ch{-i\frac{\vt{k}_\bot^2 + m^2}{2\kappa}v}\eqend{,}
\end{equation}
where $\kappa > 0$, are normalized as
\begin{equation}
 (g_{\kappa',\vt{k}_\bot'},g_{\kappa,\vt{k}_\bot})_\textrm{KG}  = 
\delta(\kappa - \kappa')\delta^{(n - 2)}(\vt{k}_\bot - \vt{k}_\bot')\eqend{.}
\end{equation}

We expand the Klein-Gordon field operator in terms of the modes $g_{\kappa,\mathbf{k}_\bot}(x)$ in Eq.~(\ref{eq:light_cone_modes_normalized}) as
\begin{eqnarray}
 \phi(x) & =&  \int_0^\infty d\kappa \int_{\mathbb{R}^{n - 2}}d^{n - 2}\textbf{k}_\bot \nonumber \\
&& \quad \times \left[a_{\kappa, \vt{k}_\bot}g_{\kappa, \vt{k}_\bot}(x)
 + a_{\kappa, \vt{k}_\bot}^\dagger g_{\kappa, \vt{k}_\bot}^*(x)\right]\eqend{.} \label{eq:KG_field_operator_expansion}
\end{eqnarray}
By imposing the commutation relations
\begin{equation}\label{eq:commutator_aa}
 [a_{\kappa, \vt{k}_\bot},a_{\kappa', \vt{k}_\bot'}] = 0\eqend{,}
\end{equation}
and
\begin{equation}\label{eq:commutator_aa_dagger}
 [a_{\kappa, \vt{k}_\bot},a_{\kappa', \vt{k}_\bot'}^\dagger] = \delta(\kappa - \kappa')\delta^{(n - 2)}(\vt{k}_\bot - \vt{k}_\bot')\eqend{.}
\end{equation}
We find the equal-time commutator
\begin{equation}\label{eq:commutator_phi_pi}
 [\phi(x),\del_{u'}\phi(x')]|_{v = v'} = \frac{i}{2}\delta(u - u')\delta^{(n - 2)}(\vt{x}_\bot - \vt{x}_\bot')\eqend{,}
\end{equation}
which is the standard equal-time commutator written differently.
The factor $1/2$ on the right-hand side of Eq.~(\ref{eq:commutator_phi_pi}) can be interpreted as coming from the fact that,
in the light-cone coordinates, the field $\phi$ and its canonical conjugate momentum are not
independent~\cite{neville_rohtlicj_nc_1971, sundermeyer_book}.

In the vacuum state $|0\rangle$ defined by $a_{\kappa, \vt{k}_\bot}|0\rangle = 0$ for all $\kappa \in \mathbb{R}^+$ and 
$\vt{k}_\bot \in \mathbb{R}^{n - 2}$, the associated Wightman two-point functions are given by
\begin{splitequation}
 G^\pm(x,x') & = \frac{1}{(2\pi)^{n - 1}}\int_{\mathbb{R}^{n - 1}}d\kappa d^{n - 2}\mathbf{k}_\bot \tilde{G}^\pm(v - v',\kappa,\vt{k}_\bot)\\
 &\qquad \qquad \qquad \qquad \times e^{i[\vt{k}_\bot \cdot (\vt{x}_\bot - \vt{x}_\bot') - \kappa (u - u')]},
\end{splitequation}
with
\begin{equation}
 \tilde{G}^\pm(v,\kappa,\vt{k}_\bot) \equiv \frac{\Theta(\pm \kappa)}{2|\kappa|} \exp\ch{-i\frac{\vt{k}_\bot^2 + m^2}{2\kappa}v}.
\end{equation}
Since the $v$-coordinate increases monotonically toward the future, the Feynman and Dyson propagators are given by
\begin{equation}\label{eq:Feynman_propagator}
 G^\textrm{F}(x,x') = \Theta(v - v')G^+(x,x') + \Theta(v' - v)G^-(x,x')\eqend{,}
\end{equation}
and
\begin{equation}
 G^\textrm{D}(x,x') = \Theta(v - v')G^-(x,x') + \Theta(v' - v)G^+(x,x')\eqend{,}
\end{equation}
respectively. Note that Eq.~(\ref{eq:Feynman_propagator}) corresponds to the form of the Feynman propagator given by Eq.~(\ref{eq:Feynman_propagator_lightcone}).

\bibliography{hl_v6}

\end{document}